\documentclass[sigconf, authorversion, nonacm]{acmart}

\acmConference[WSESE@ICSE 2024]{46th International Conference on Software Engineering}{April 2024}{Lisbon, Portugal}

\usepackage{algorithmic}
\usepackage{graphicx}
\usepackage{textcomp}
\usepackage{xcolor}
\usepackage[most]{tcolorbox}
\usepackage{printlen}
\usepackage{pgfplots}
\usepackage{float}
\usepackage{pgfplots}
\usepackage[font={small}]{caption}
\usepackage{multicol}
\usepackage{subcaption}
\PassOptionsToPackage{hyphens}{url}
\usepackage{cleveref}
\usepackage{listings}
\usepackage{tabularx}

\title{Lessons Learned from Mining the Hugging Face Repository}

\author{Joel Castaño}
\affiliation{%
  \institution{Universitat Politècnica de Catalunya}
  \city{}
  \country{}}
\email{joel.castano@upc.edu}

\author{\normalfont Silverio Martínez-Fernández}
\affiliation{%
  \institution{
  Universitat Politècnica de Catalunya}
  \city{}
  \country{}
}
\email{silverio.martinez@upc.edu}

\author{Xavier Franch}
\affiliation{%
 \institution{Universitat Politècnica de Catalunya}
 \city{}
 \country{}
}
\email{xavier.franch@upc.edu}

\date{October 2023}

\begin{abstract}
The rapidly evolving fields of Machine Learning (ML) and Artificial Intelligence have witnessed the emergence of platforms like Hugging Face (HF) as central hubs for model development and sharing. This experience report synthesizes insights from two comprehensive studies conducted on HF, focusing on carbon emissions and the evolutionary and maintenance aspects of ML models. Our objective is to provide a practical guide for future researchers embarking on mining software repository studies within the HF ecosystem to enhance the quality of these studies. We delve into the intricacies of the replication package used in our studies, highlighting the pivotal tools and methodologies that facilitated our analysis. Furthermore, we propose a nuanced stratified sampling strategy tailored for the diverse HF Hub dataset, ensuring a representative and comprehensive analytical approach. The report also introduces preliminary guidelines, transitioning from repository mining to cohort studies, to establish causality in repository mining studies, particularly within the ML model of HF context. This transition is inspired by existing frameworks and is adapted to suit the unique characteristics of the HF model ecosystem. Our report serves as a guiding framework for researchers, contributing to the responsible and sustainable advancement of ML, and fostering a deeper understanding of the broader implications of ML models. 
\end{abstract}



\def\BibTeX{{\rm B\kern-.05em{\sc i\kern-.025em b}\kern-.08em
    T\kern-.1667em\lower.7ex\hbox{E}\kern-.125emX}}

\usepackage{todonotes}

\usepackage{csquotes}

\pgfplotsset{compat=1.18}

\usepackage{array}

\begin{document}

\maketitle

\section{Introduction}

In the fast-evolving fields of Machine Learning (ML) and Artificial Intelligence (AI), resources such as Hugging Face (HF) and its Hub \cite{HuggingFaceInc.2023} have become pivotal. HF, particularly renowned for its facilitation of Large Language Models and generative AI, combines the technical intricacies of ML models with the collaborative dynamics of a global community of researchers and developers. This evolving ecosystem not only drives innovation and knowledge sharing but also underscores the need to understand the behaviors, methodologies, and impacts of these models.

This experience report aims to empower future researchers with practical knowledge and insights for conducting mining software repository studies on HF. It not only synthesizes lessons learned from our studies, detailed in \cite{castano2023exploring} and \cite{castano2023analyzing}, but also incorporates insights from other existing repository mining studies on the HF platform, further depicted in Section \ref{sec:background}. These collective experiences and findings have been essential in shaping our understanding and framing the context of our report. By sharing our experiences and methodologies, this report serves as a guiding framework for researchers and practitioners embarking on similar analytical explorations in the field of ML.

We reflect on three key lessons learned addressing methodological issues for mining HF. Firstly, we acknowledge the need to offer practical guidance's to future practitioners on HF mining. To address this, we delve into a detailed review of the replication package used in our studies, focusing on the pivotal tools, datasets, and methodologies. This examination is aimed at demystifying the components of the package and facilitating its more accessible and efficient use by future researchers. Secondly, to provide a comprehensive understanding of the HF Hub's landscape, it is crucial to recognize that focusing solely on the most popular models may not always yield the most representative or insightful results (e.g., user may want to consider representation on various categories). Consequently, we propose adopting an optional stratified sampling strategy, designed for researchers analyzing data from HF, taking into account the wide array of models and data available. Finally, we discuss a methodological shift from repository mining to cohort studies. This shift is primarily motivated by the need to establish a more rigorous approach than traditional correlation-based studies, aiming for deeper insights and a closer approximation to causality, particularly in the context of ML models of HF.  This is inspired by \cite{saarimaki2020cohort} and \cite{saarimaki2023does}, which have been illustrating such methodologies. Our adoption of this methodology offers a contribution to the domain of empirical software engineering.

The structure of our paper is organized into three distinct sections. Initially, Section \ref{sec:background} sets the stage by presenting the necessary context and introducing key concepts integral to our studies. This is followed by Section \ref{sec:lessons_learned} where we explicitly state the methodological insights derived from our analyses. Finally, Section \ref{sec:discussions_outlook} is dedicated to discussing the implications and outlooks of our work, where we delve into the broader impact and potential future directions that stem from our findings and experiences.

\textbf{Data Availability Statement}: The replication package, along with the resulting dataset, is made available on Zenodo \cite{castano_fernandez_zenodo}.

\section{BACKGROUND AND RELATED WORK}\label{sec:background}

\subsection{Repository Mining}

Repository mining, a systematic approach for quantitatively analyzing datasets from platforms hosting structured or semi-structured text \cite{vidoni2022systematic}, is crucial for understanding the HF platform in our study. By using repository mining, we accessed and analyzed data, including model metadata, commit messages, and discussion threads, to gain insights into community engagement, model evolution, and maintenance practices within the HF ecosystem.

Our study's methodology aligns with repository mining by: (i) justifying its application for our research problem; (ii) defining analysis units and describing the HF Hub API as our data source; (iii) outlining the selection process, criteria, and validation methods for the data; (iv) detailing data preprocessing steps for integrity and relevance; and (v) acknowledging potential threats to external validity from our data source selection.

We also ensure our approach's alignment with repository mining's desirable attributes by providing supplemental materials, like datasets and processing code, and suggesting future research avenues that could utilize our dataset and findings.

\subsection{Repository Mining in HF}

Repository mining studies on HF have collectively enriched our understanding of the platform's landscape in ML model development. Kathitar et al. \cite{Kathikar2023} conducted a vital security analysis, uncovering a presence of high-severity vulnerabilities within HF's GitHub-linked repositories. This study underscores the complexities of securing ML models in open-source ecosystems. In our own work, \cite{castano2023exploring}, the focus was on the environmental impact of ML models on HF, particularly their carbon footprint, highlighting the need for sustainable development practices in the ML community.

Additionally, \citet{AIT2024103079} introduced \textit{HFCommunity}, a tool that facilitates the empirical analysis of ML projects on HF collecting and integrating data of HF, emphasizing the platform's growing role as a hub for collaborative development. Meanwhile, Jiang et al. \cite{Jiang2023} delved into the practices and challenges of pre-trained model reuse in HF, offering a perspective on model maintenance and dependency management in the ML ecosystem. Our subsequent study, \cite{castano2023analyzing}, further explored the community engagement, evolution, and maintenance of over 380,000 models on HF, providing a comprehensive view of model development trends and maintenance practices. The work by Pepe et al. \cite{pepe2023fairness}, delves into crucial aspects of fairness, bias, and legal issues associated with pre-trained models, adding an important dimension to the understanding of these models in practice.  Lastly, \citet{jiang2023exploring} examined the naming conventions and defects of pre-trained models in HF, shedding light on the research-to-practice pipeline in the PTM ecosystem.

These studies collectively pave the way for the usage of the proposed replication package, sampling strategy, and cohort study, offering a structured approach to further investigate Hugging Face.

\subsection{Sampling and Stratified Sampling}

Sampling is key in empirical software engineering studies \cite{de2015investigating}. It is a statistical method used in research to select a subset of data or observations from a larger dataset or population \cite{cochran1977sampling}. The purpose of sampling is to draw conclusions about the entire dataset or population by examining only a part of it, thereby making the research process more manageable and cost-effective.

Stratified sampling, a specific type of sampling method, involves dividing the population into distinct subgroups or strata that share similar characteristics. Each stratum is then sampled as an independent subpopulation, out of which individual elements are randomly selected. The key advantage of stratified sampling lies in its efficiency and the increased accuracy it provides, especially when certain strata in the population are known to vary. By ensuring that each subgroup is adequately represented, stratified sampling can produce more representative and reliable results than simple random sampling.

In the context of our experience report, we propose a stratified sampling strategy to analyze the ML models on the HF platform. This approach enables us to obtain a representative overview of the different types of models present on the platform, ensuring that our conclusions and insights are well-grounded and reflective of the diverse nature of the models hosted on HF.

\subsection{Cohort Studies and Their Adaptation in Mining Software Repositories}

A cohort study is a type of observational study commonly used in medical and social science research \cite{hennekens1987epidemiology}. It involves following a group of individuals (a cohort) over a period of time to observe and record various outcomes. In this type of study, the cohort is selected based on certain shared characteristics or experiences, and data is collected at various intervals to analyze how these characteristics influence specific outcomes over time.

In the context of mining software repositories, there are still many methodological challenges \cite{ayala2021use} (e.g., the difficulty in controlling and manipulating the data used in studies). A cohort study can be adapted to examine the evolution and impacts of software development practices, tools, or technologies within a specific community or platform. This approach allows researchers to track changes and developments over time, offering insights into trends, patterns, and potential causal relationships.

Cohort studies have been explored in software engineering, such as Saarimäki et al.~\cite{saarimaki2020cohort}. In the context of HF studies, Saarimäki et al.~\cite{saarimaki2023does} also provided a detailed framework on how cohort studies can be effectively adapted for mining software repository studies. Key aspects include:

\begin{itemize}
    \item Identification of cohorts based on shared characteristics within the software repository data, such as programming languages used, types of projects, or development practices.
    \item Longitudinal tracking of these cohorts to observe how changes in software development practices or technologies influence various outcomes, such as software quality, developer productivity, or project success.
    \item Use of statistical methods to analyze the data and draw conclusions about trends and patterns observed within the cohorts. This may include techniques like Spearman’s non-parametric correlation or the Mann-Whitney U test to handle non-normally distributed data.
\end{itemize}

We apply these principles to propose a cohort study for the HF model ecosystem. By treating each model or group of models as a cohort, we aim to determine causality by design on how different factors, such as model size or training tasks, influence key outcomes like carbon emissions and model efficiency.

\section{HF Mining: Methodological Insights}
\label{sec:lessons_learned}

\subsection{Replication Package}

\begin{figure}[h]
\centering
\includegraphics[width=0.8\columnwidth]{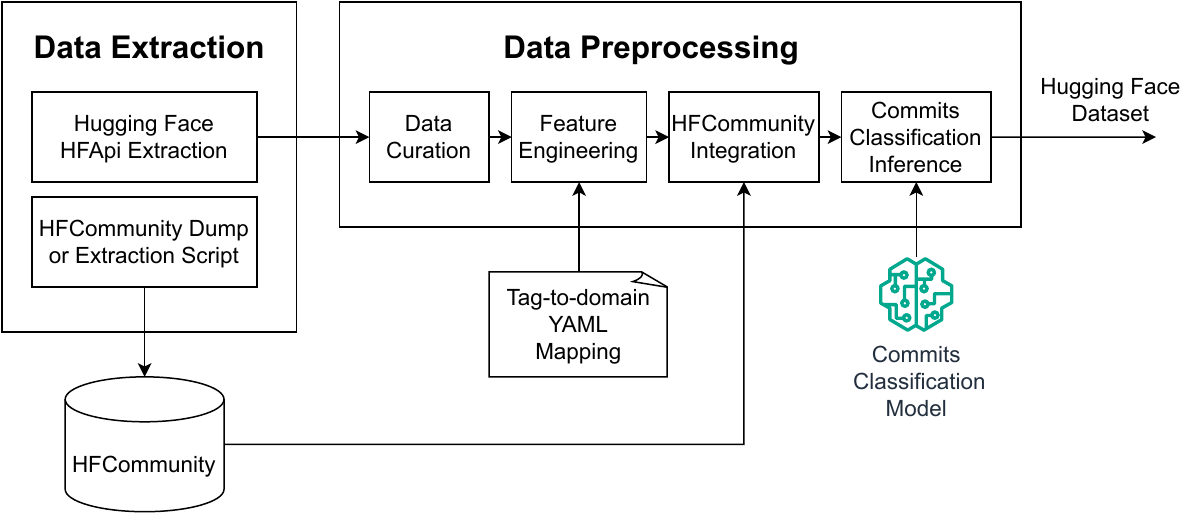}
\caption{Replication package diagram}
\label{replication_package_diagram}
\end{figure}

In this section, we detail a replication package for the HF Hub, which amalgamates the methodologies from our two prior studies \cite{castano2023exploring} \cite{castano2023analyzing}. We have included a diagram of the replication package in Figure \ref{replication_package_diagram} for a clearer understanding. Our approach aims to provide a robust and versatile dataset, serving as a valuable resource for future research endeavors wanting to analyze the HF community, particularly focusing on aspects such as carbon emissions, model evolution, and maintenance practices. 

\subsubsection{\textbf{Data Collection}}

Our data collection process leverages the HfApi class of the HF Hub API, a Python wrapper for efficient interaction within the HF platform. This wrapper allows us to retrieve metadata of every model that has been uploaded to HF.

After retrieving the list of models, our pipeline automates the process of extracting information. This includes using regular expressions to identify and gather evaluation metrics such as accuracy, F1 and Rouge1, among others. Each entry in the dataframe created by our pipeline represents a model with various attributes. We collect a wide range of model attributes:

\begin{itemize}
    \item \textbf{General model information}: Attributes such as the model size, popularity indicators (downloads and likes), creation dates, associated tags reflecting their features or domains (e.g., PyTorch, Transformer, \ldots), among others.
    \item \textbf{Training details}: Including datasets sizes and evaluation metrics (e.g., accuracy or F1) if reported.
    \item \textbf{Maintenance and evolutionary information}: Detailed data related to the commit history and discussions of models.
    \item \textbf{Carbon emissions and key related-context}: The reported training CO$_2$e emissions, the hardware used, the geographical location, among others.
    \item \textbf{Model card text raw data}: The raw model card written to describe the model.
\end{itemize}

For more comprehensive data regarding commit histories files editing, the \textit{HFCommunity} dataset is employed \cite{Ait2023}. HFCommunity offers an offline up-to-date relational database on HF data. The integration of data from the \textit{HFCommunity} dataset, which uses the PyDriller framework, is necessary to provide the list of files edited in each commit. This requires setting up a local MariaDB database dumping the database provided in their website. 

Through its versioning system, HF maintains historical data such as changes in model files (e.g., \textit{pytorch\_model.bin}), updates to model cards, and commit message histories. However, some data, like time-specific download counts, user interactions, and detailed model performance evolution, are not preserved in this system. To obtain these dynamic insights, crucial for analyzing model popularity and engagement, we need to employ incremental crawling and periodic data collection.

Moreover, the extraction process from HF, particularly through API calls, is optimized for efficiency via parallelization. Using multiple \textit{jobs} with concurrent threads significantly reduces the data collection time, a crucial aspect given the vast scale of the dataset (over 380,000 models as of November 6, 2023). This approach, however, needs careful management to avoid API request limitations and ensure stable data retrieval.

\subsubsection{\textbf{Data Preprocessing}}

The data preprocessing stage is a crucial step in transforming the raw data collected from the HF Hub API and the HFCommunity dataset into a structured, clean, and analysis-ready format. This stage involves several key processes to handle the diverse nature of the data, ensure consistency, and enhance the data's utility research objectives.

\begin{itemize}

\item \textbf{Feature Engineering:} We perform feature engineering for enhanced analysis (e.g., creating variables like \textit{domain} from tag-to-domain mapping)

\item \textbf{Tag Filtering and Manipulation:} Tags are filtered to remove non-specific or auxiliary tags (e.g., languages, 'license', 'dataset'). We also one-hot encode the remaining tags for analytical purposes.

\item \textbf{Data Enrichment and Cleaning:} The dataset undergoes cleaning to manage missing values and harmonize data formats. A subset of the data, especially regarding CO$_2$e emissions, is particularly curated for accuracy and completeness.

\item \textbf{Handling of unstructured data: } We need to deal dealing with the high volume of unstructured data, particularly from commit and discussion storage. This needs careful handling to avoid encoding issues and ensure that the data was transformed into a usable dictionary format.

\item \textbf{Commit Data Classification:} Using a fine-tuned neural network, in particular a DistilBERT model based on the proposal from Sarwar et al. \cite{sarwar2020multi}, we categorize commit messages into Corrective, Perfective, and Adaptive, based on Swanson's software maintenance categories \cite{swanson1976dimensions}.

\end{itemize}

The final dataset comprises over 380,000 model entries from HF, enriched with detailed development and maintenance history. This dataset forms the basis of our study and offers a versatile resource for future research.

\subsubsection{\textbf{Automation of the Data Update}}

The automation process primarily involves the following steps:

\begin{itemize}
    \item \textbf{Data Extraction}: The use of the HF Hub API with the HfApi class lends itself well to automation. Scripts can be adapted to automatically query the API at scheduled intervals.
    \item \textbf{Integrating Data from HFCommunity}: Automation of data integration from the HFCommunity dataset is achievable setting up a local instance of the MariaDB database containing HFCommunity data and scheduling regular updates. Scripts can be developed to automatically fetch and integrate this data into the main dataset.
    \item \textbf{Data Preprocessing and Classification}: The preprocessing steps, including data cleaning, harmonization, and commit classification, can be automated through the scheduling of the script without major modifications. Automated scripts can also handle tag-to-domain mapping and other preprocessing tasks.
\end{itemize}

\paragraph{\textbf{Considerations and Challenges}}

While the theoretical framework for automation is established, there are practical challenges that need to be addressed:

\begin{itemize}
    \item \textbf{Resource-Intensive Processes}: Both data extraction and preprocessing are resource-intensive. Efficient handling of API calls requires parallel processing to reduce time. Preprocessing, especially the neural network-based classification of commit data, demands significant computational power, ideally with a GPU setup for efficient processing.
    \item \textbf{HFCommunity Data Integration}: Accessing HFCommunity data necessitates a local database configuration. The initial setup and subsequent updates from HFCommunity can be resource and time-intensive.
    \item \textbf{Maintaining Data Integrity and Consistency}: Automation must ensure that the integrity and consistency of the data are maintained. This includes handling updates or changes in the HF API or HFCommunity dataset structure.
    \item \textbf{Handling API Limitations}: There is a need to manage API limitations, such as the number of requests per second. Use of API tokens and error handling mechanisms must be incorporated to mitigate the risks of request rejections.
    \item \textbf{Low Quality Artifacts}: Our data preprocessing currently employs basic filtering to ensure artifact relevance and quality on HF. We acknowledge the need for more advanced quality control measures in future work to thoroughly assess and filter artifacts, enhancing their reliability and integrity.
\end{itemize}

\paragraph{\textbf{Feasibility and Implementation}}

Therefore, implementing the automation of the data update process is feasible but requires careful planning and resource allocation:

\begin{itemize}
    \item \textbf{Scalability}: The system should be scalable to handle the increasing volume of data from HF and HFCommunity.
    \item \textbf{Parallel Processing}: Implementing parallel processing for API calls and data processing is essential for time efficiency.
    \item \textbf{Robust Computational Resources}: Ensuring access to robust computational resources, including powerful GPUs and database setups, is critical for handling large-scale data processing and neural network computations.
    \item \textbf{Continuous Monitoring and Updating}: Automated systems require continuous monitoring to address any issues arising from API changes or data inconsistencies.
\end{itemize}

\subsection{Sampling}

Given the diverse range of attributes in HF,  representing a population that includes all ML models hosted on the platform across various domains and characteristics, a stratified sampling method is proposed to ensure representativeness across various dimensions of the data.  The sampling approach is further depicted in Figure \ref{stratification_diagram}. This method involves dividing the dataset into distinct strata based on specific attributes and then randomly sampling from each stratum. The key stratification criteria are as follows:

\begin{figure}[h]
\centering
\includegraphics[width=0.8\columnwidth]{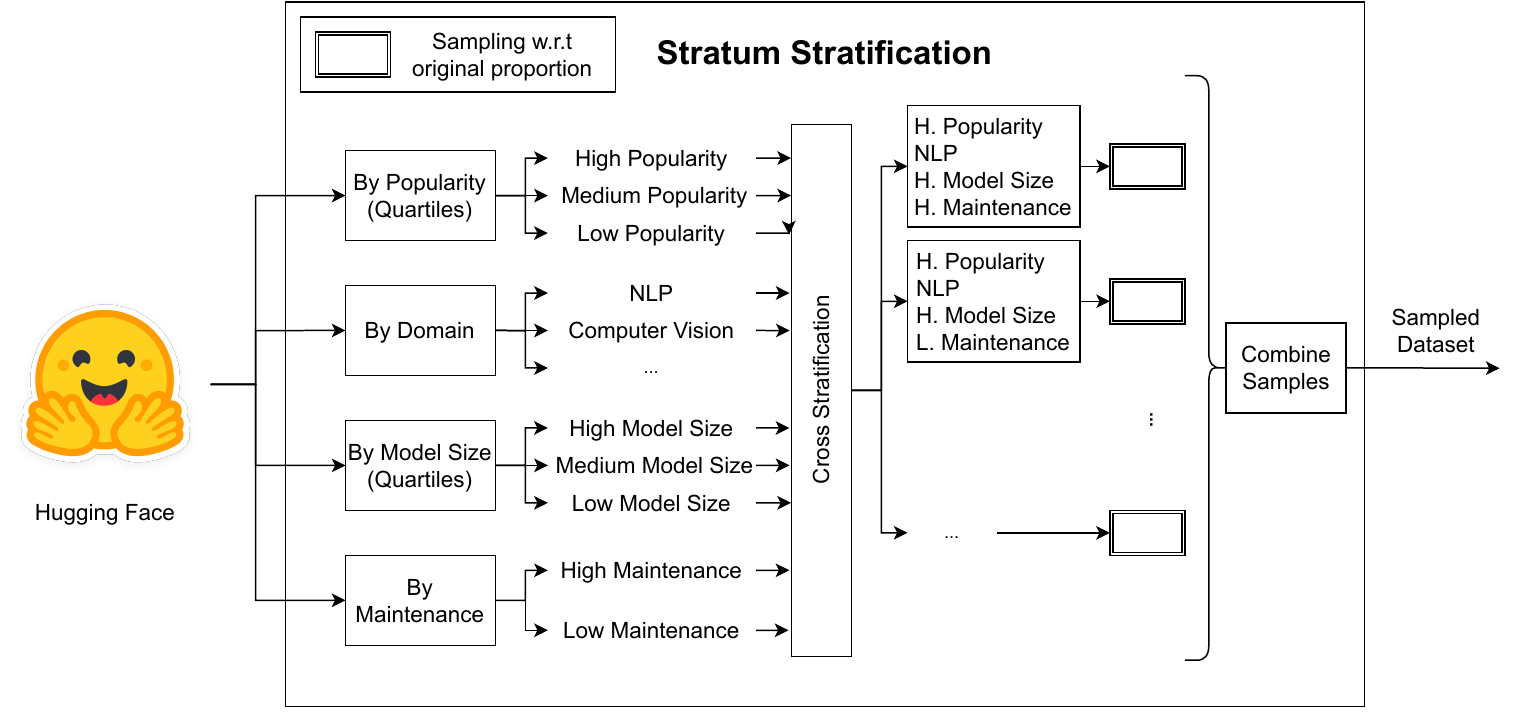}
\caption{Stratification sampling diagram}
\label{stratification_diagram}
\end{figure}

\begin{itemize}
    \item \textbf{Model Domain:} The dataset is categorized into distinct domains such as NLP, Computer Vision, Multimodal, Audio, and Reinforcement Learning, reflecting the classification system proposed by HF \cite{huggingfaceModelsHugging}. This categorization helps in examining domain-specific trends and practices.

    \item \textbf{Model File Size:} The size attribute is divided into quartiles, allowing for a comparative analysis across models of varying sizes and complexities. This helps in understanding if model size correlates with factors like maintenance frequency, popularity, or carbon efficiency.

    \item \textbf{Popularity:} Popularity metric divided into quartiles, computed as the sum of the normalized likes and downloads, is used to create strata representing different levels of community engagement and recognition.
\end{itemize}

To ensure disjoint strata and avoid duplicates, the final strata are constructed as the 'cross product' of these key criteria. For example, combining model domain and model file size would yield strata such as NLP-High Model Size, NLP-Low Model Size, Computer Vision-High Model Size, etc. This granular approach, facilitated by the abundance of models in HF, allows for a nuanced sampling that captures unique intersections of characteristics.

\textit{Optional Further Stratification:} While the main criteria provide a broad and general stratification, further granularity can be achieved on demand by incorporating additional criteria:
\begin{itemize}
\item \textbf{Maintenance Classification:} Classifying models as High or Low Maintenance, as detailed in \cite{castano2023analyzing}. This classification provides insights into the ongoing effort invested in model upkeep and improvement.
\item \textbf{Carbon Efficiency Classification:} Categorizing models based on carbon efficiency (A to E) as presented in \cite{castano2023exploring}. 
\item \textbf{Library usage:} Main library usage (e.g., \textit{PyTorch}).
\end{itemize}
These optional criteria can be applied to already formed strata for deeper insights, especially when specific aspects such as maintenance practices or environmental impact are of interest. The proposed stratification is flexible enough to be easily adapted on to attributes for various research needs within the HF ecosystem. This ensures the methodology remains relevant across studies.

\subsubsection{\textbf{Sampling Procedure}}
The sampling process involves the following steps:

\begin{enumerate}
    \item \textbf{Strata Formation:} For each criterion, the dataset is divided into respective strata. We form disjoint strata by combining attributes from each key criterion. For example, the model domain, model file and popularity creates separate strata for NLP-High Popularity-High Model Size,  NLP-High Popularity-Low Model Size, etc.

    \item \textbf{Sample Size Determination:} Initially, the total sample size required for the study is determined considering research needs and ensuring robust statistical analysis (e.g., through statistical formulas as proposed in \cite{kitchenham2002principles}). Once the total sample size is established, the sample size for each stratum is then determined based on its proportion in the overall dataset to maintain representativeness across the various strata. 

    \item \textbf{Random Sampling:} Within each stratum, a random sample of models is selected. This step ensures that each model within a stratum has an equal chance of being selected.

    \item \textbf{Composite Sample Construction:} The final sample is a composite of the randomly selected models from each stratum, providing a representative cross-section of the dataset.
\end{enumerate}

\subsubsection{\textbf{Data Analysis Techniques for Stratified Samples}}
In addition, to effectively analyze stratified samples, advanced techniques of meta-analysis can also be employed \cite{hedges2014statistical}. These techniques allow for the generalization of correlation measures and other statistical methods across different strata, ensuring a comprehensive understanding of the dataset. For example, adapted versions of Pearson correlation and other relevant measures could be used, which cater to the nuances of stratified sampling.

\subsubsection{\textbf{Expected Outcomes}}
This stratified sampling approach is expected to yield a representative subset of the HF dataset, facilitating an in-depth analysis of the various aspects under investigation in a  study, such as model maintenance practices, carbon efficiency, and popularity trends across different model domains and sizes.

\subsection{Cohort Study for HF Ecosystem}

In this section, we present preliminary guidelines on how to conduct cohort studies in the context of the HF model ecosystem. This approach adapts traditional epidemiological cohort study methods to the unique environment of software repository mining. Next we define the main takeovers future researchers should take into account for addressing a cohort study on HF.

\subsubsection{Defining Research Objectives}
Formulating research objectives is a critical step in a cohort study. Objectives should focus on understanding the impact of certain factors on model development, usage, and maintenance within the repository.

\subsubsection{Cohort Identification and Stratification}
The first step in a cohort study is to identify and define the cohorts. Cohorts within the HF ecosystem can be based on various shared characteristics of the ML models. Cohort identification will align with the stratified sampling methodology presented earlier, ensuring representativeness across different dimensions of the dataset.

\begin{itemize}
    \item \textbf{Defining Cohorts:} Criteria for defining cohorts may include model architecture, training tasks, dataset sizes, programming languages, and usage metrics.
    \item \textbf{Cohort Sampling:} Sample size and selection within each cohort should reflect the sampling approach, providing a balanced representation across various model attributes.
\end{itemize}

\subsubsection{Data Collection and Longitudinal Tracking}
Data collection and longitudinal tracking involve gathering and analyzing data over a specified period. A key feature of cohort studies is their longitudinal nature. In the HF context, this involves tracking changes in model development practices, maintenance activities, and community engagement over time. Regular data collection intervals should be established, aligning them with significant events in the HF ecosystem, such as major updates or community initiatives.

By accessing multiple versions of models through their commit histories, researchers can conduct a comprehensive longitudinal analysis. This approach allows for observing the evolution of models over time, providing critical insights into their development and maintenance lifecycle. Cohorts should be monitored over a set period, with data collection intervals determined based on model evolution dynamics and research objectives.

\subsubsection{Statistical Analysis and Interpretation}
Analyzing the collected data is essential for drawing meaningful conclusions from the cohort study.

\begin{itemize}
    \item \textbf{Analytical Techniques:} Statistical methods suitable for longitudinal and non-normally distributed data should be employed, including Spearman’s correlation and Mann-Whitney U tests.
    \item \textbf{Handling Confounders:} Identify potential confounders unique to the HF ecosystem, such as the effect of community events (e.g., competitions, collaborative projects), changes in HF platform features, or the introduction of new technologies and model architectures. Strategies such as stratification and matching, aligned with the sampling methodology, should be implemented to ensure comparability between different cohorts.
\end{itemize}

\subsubsection{Observational Nature of the Study}

In the HF ecosystem, the observational nature of cohort studies could focus on:

\begin{itemize}
\item Systematic tracking of changes in model development practices, including updates in code, model architecture modifications, and alterations in documentation (model cards).
\item Monitoring community engagement metrics such as likes, downloads, and discussion activities around different models to gauge popularity and user interest.
\item Observing how models are used across different domains and tasks, paying special attention to reusability and adaptability of models in various contexts.
\end{itemize}

\subsubsection{Implications and Reporting}
The final stage entails interpreting results and reporting findings, focusing on the HF platform's unique dynamics, including collaborative model development and user diversity. Reporting should be transparent, reproducible, and utilize visual aids for clarity.

This guideline is specifically adapted to the dynamics of the HF platform and should be refined iteratively based on emerging trends and feedback from the community.

\section{Discussions and Outlook}\label{sec:discussions_outlook}

Our experience report, grounded in studies conducted on HF, unveils insights and practical strategies for future researchers in the realm of ML.

\textbf{Importance of Replication and Accessibility:} The exploration of the replication package underlines the importance of transparency and accessibility in research. We empower researchers to build upon existing work through the presented replication package, fostering a collaborative and cumulative research environment.

\textbf{Stratified Sampling Strategy:} The proposed stratified sampling strategy serves the necessity of nuanced and methodical approaches in handling a diverse and voluminous dataset. This strategy enhances the representativeness of samples and ensures that findings are reflective of the multifaceted nature of the HF ecosystem. 

\textbf{Cohort Studies in Software Repository Mining:} The preliminary guidelines for adapting cohort studies to the context of software repository mining arguably set a methodological milestone in this area. This approach is anticipated to enable the longitudinal study of ML models, offering initial steps towards determining causality within repository mining studies. We welcome discussions and collaborations within the MSR community to refine and enhance these guidelines further.

\textbf{Standardization Challenges in Reporting:} The replication package construction revealed a notable lack of standardization in the reporting of HF models. This inconsistency poses challenges for researchers aiming to conduct comprehensive and comparative studies. Addressing this issue is crucial for improving the reliability and utility of future research on the HF platform.

As we look ahead, the methodologies and lessons learned from our work open pathways for more profound and extensive research in HF. Future research and considerations focus on:

\textbf{Enhancing Cohort Study Frameworks:} Continued development and refinement of cohort study methodologies tailored for ML platforms like HF. This includes integrating feedback from the MSR community and adapting the framework to the dynamic nature of ML model development and maintenance.

\textbf{Promoting Reporting Standardization:} Advocating for and contributing to the standardization of model reporting on platforms like HF. This could involve developing guidelines that aid in the uniform reporting of model characteristics and metrics.

\textbf{Adaptability to Other Platforms and Repositories:} While our study is focused on the HF platform, the strategies and lessons learned can be adapted to other ML platforms and repositories (PyTorch Hub, TensorFlow Hub). 

\textbf{Continual Learning and Iterative Improvement:} The field of ML is characterized by rapid change and innovation. As such, continual learning and iterative improvement of methodologies will be essential. Our report provides a foundation, but it is imperative that future research remains flexible and responsive to new developments and emerging trends in the field.

In conclusion, our experience report provides valuable insights into the current state of ML models on the HF platform and also opens doors for continued innovation and exploration in the field of ML. By sharing our methodologies and findings, we aim to foster a more collaborative, transparent, and rigorous research environment in HF.

\section*{ACKNOWLEDGMENTS}
This work is supported by the project TED2021-130923B-I00, funded by MCIN/AEI/10.13039/501100011033 and the European Union Next Generation EU/PRTR.

\bibliographystyle{IEEEtranN}
\bibliography{References}

\begin{thebibliography}{22}
\providecommand{\natexlab}[1]{#1}
\providecommand{\url}[1]{#1}
\csname url@samestyle\endcsname
\providecommand{\newblock}{\relax}
\providecommand{\bibinfo}[2]{#2}
\providecommand{\BIBentrySTDinterwordspacing}{\spaceskip=0pt\relax}
\providecommand{\BIBentryALTinterwordstretchfactor}{4}
\providecommand{\BIBentryALTinterwordspacing}{\spaceskip=\fontdimen2\font plus
\BIBentryALTinterwordstretchfactor\fontdimen3\font minus \fontdimen4\font\relax}
\providecommand{\BIBforeignlanguage}[2]{{%
\expandafter\ifx\csname l@#1\endcsname\relax
\typeout{** WARNING: IEEEtranN.bst: No hyphenation pattern has been}%
\typeout{** loaded for the language `#1'. Using the pattern for}%
\typeout{** the default language instead.}%
\else
\language=\csname l@#1\endcsname
\fi
#2}}
\providecommand{\BIBdecl}{\relax}
\BIBdecl

\bibitem[{Hugging Face Inc.}(2023)]{HuggingFaceInc.2023}
{Hugging Face Inc.}, ``Hugging {{Face Hub Documentation}},'' \url{https://huggingface.co/docs/hub/index}, 2023.

\bibitem[Casta{\~n}o et~al.(2023)Casta{\~n}o, Mart{\'\i}nez-Fern{\'a}ndez, Franch, and Bogner]{castano2023exploring}
J.~Casta{\~n}o, S.~Mart{\'\i}nez-Fern{\'a}ndez, X.~Franch, and J.~Bogner, ``Exploring the carbon footprint of hugging face's ml models: A repository mining study,'' \emph{arXiv preprint arXiv:2305.11164}, 2023.

\bibitem[\vspace{0mm} Casta{\~n}o et~al.(2023)\vspace{0mm} Casta{\~n}o, Mart{\'\i}nez-Fern{\'a}ndez, Franch, and Bogner]{castano2023analyzing}
J.~\vspace{0mm} Casta{\~n}o, S.~Mart{\'\i}nez-Fern{\'a}ndez, X.~Franch, and J.~Bogner, ``Analyzing the evolution and maintenance of ml models on hugging face,'' \emph{arXiv preprint arXiv:2311.13380}, 2023.

\bibitem[Saarim{\"a}ki et~al.(2020)Saarim{\"a}ki, Lenarduzzi, Vegas, Juristo, and Taibi]{saarimaki2020cohort}
N.~Saarim{\"a}ki, V.~Lenarduzzi, S.~Vegas, N.~Juristo, and D.~Taibi, ``Cohort studies in software engineering: A vision of the future,'' in \emph{Proceedings of the 14th ACM/IEEE International Symposium on Empirical Software Engineering and Measurement (ESEM)}, 2020, pp. 1--6.

\bibitem[Saarimaki et~al.(2023)Saarimaki, Manero, Juristo, Taibi, Lenarduzzi, et~al.]{saarimaki2023does}
N.~Saarimaki, M.~R. Manero, N.~Juristo, D.~Taibi, V.~Lenarduzzi \emph{et~al.}, ``Does microservices adoption impact the development velocity? a cohort study. a registered report,'' \emph{arXiv preprint arXiv:2306.02034}, 2023.

\bibitem[Castaño~Fernández et~al.(2023)Castaño~Fernández, Martínez-Fernández, and Franch]{castano_fernandez_zenodo}
\BIBentryALTinterwordspacing
J.~Castaño~Fernández, S.~Martínez-Fernández, and X.~Franch, ``{Replication Package for 'Lessons Learned from Mining the Hugging Face Repository'},'' Dec. 2023. [Online]. Available: \url{https://doi.org/10.5281/zenodo.10292130}
\BIBentrySTDinterwordspacing

\bibitem[Vidoni(2022)]{vidoni2022systematic}
M.~Vidoni, ``A systematic process for mining software repositories: Results from a systematic literature review,'' \emph{Information and Software Technology}, vol. 144, p. 106791, 2022.

\bibitem[Kathikar et~al.(2023)Kathikar, Nair, Lazarine, Sachdeva, and Samtani]{Kathikar2023}
A.~Kathikar, A.~Nair, B.~Lazarine, A.~Sachdeva, and S.~Samtani, ``Assessing the {Vulnerabilities} of the {Open}-{Source} {Artificial} {Intelligence} ({AI}) {Landscape}: {A} {Large}-{Scale} {Analysis} of the {Hugging} {Face} {Platform},'' in \emph{{IEEE} {Intelligence} and {Security} {Informatics}}.\hskip 1em plus 0.5em minus 0.4em\relax Charlotte, NC, USA: IEEE, Oct. 2023.

\bibitem[Ait et~al.(2024)Ait, {Cánovas Izquierdo}, and Cabot]{AIT2024103079}
\BIBentryALTinterwordspacing
A.~Ait, J.~L. {Cánovas Izquierdo}, and J.~Cabot, ``Hfcommunity: An extraction process and relational database to analyze hugging face hub data,'' \emph{Science of Computer Programming}, vol. 234, p. 103079, 2024. [Online]. Available: \url{https://www.sciencedirect.com/science/article/pii/S0167642324000029}
\BIBentrySTDinterwordspacing

\bibitem[Jiang et~al.(2023{\natexlab{a}})Jiang, Synovic, Hyatt, Schorlemmer, Sethi, Lu, Thiruvathukal, and Davis]{Jiang2023}
\BIBentryALTinterwordspacing
W.~Jiang, N.~Synovic, M.~Hyatt, T.~R. Schorlemmer, R.~Sethi, Y.-H. Lu, G.~K. Thiruvathukal, and J.~C. Davis, ``An {Empirical} {Study} of {Pre}-{Trained} {Model} {Reuse} in the {Hugging} {Face} {Deep} {Learning} {Model} {Registry},'' in \emph{2023 {IEEE}/{ACM} 45th {International} {Conference} on {Software} {Engineering} ({ICSE})}.\hskip 1em plus 0.5em minus 0.4em\relax Melbourne, Australia: IEEE, May 2023, pp. 2463--2475. [Online]. Available: \url{https://ieeexplore.ieee.org/document/10172757/}
\BIBentrySTDinterwordspacing

\bibitem[Pepe and Di~Penta(2023)]{pepe2023fairness}
F.~Pepe and M.~Di~Penta, ``Fairness, bias, and legal issues in pretrained models: an empirical study,'' in \emph{EMELIOT Workshop at ISSSE}, 2023.

\bibitem[Jiang et~al.(2023{\natexlab{b}})Jiang, Cheung, Thiruvathukal, and Davis]{jiang2023exploring}
W.~Jiang, C.~Cheung, G.~K. Thiruvathukal, and J.~C. Davis, ``Exploring naming conventions (and defects) of pre-trained deep learning models in hugging face and other model hubs,'' \emph{arXiv preprint arXiv:2310.01642}, 2023.

\bibitem[de~Mello et~al.(2015)de~Mello, Da~Silva, and Travassos]{de2015investigating}
R.~M. de~Mello, P.~C. Da~Silva, and G.~H. Travassos, ``Investigating probabilistic sampling approaches for large-scale surveys in software engineering,'' \emph{Journal of Software Engineering Research and Development}, vol.~3, no.~1, pp. 1--26, 2015.

\bibitem[Cochran(1977)]{cochran1977sampling}
W.~G. Cochran, \emph{Sampling techniques}.\hskip 1em plus 0.5em minus 0.4em\relax john wiley \& sons, 1977.

\bibitem[Hennekens and Buring(1987)]{hennekens1987epidemiology}
C.~H. Hennekens and J.~E. Buring, ``Epidemiology in medicine,'' in \emph{Epidemiology in medicine}, 1987, pp. 383--383.

\bibitem[Ayala et~al.(2021)Ayala, Turhan, Franch, and Juristo]{ayala2021use}
C.~Ayala, B.~Turhan, X.~Franch, and N.~Juristo, ``Use and misuse of the term “experiment” in mining software repositories research,'' \emph{IEEE Transactions on Software Engineering}, vol.~48, no.~11, pp. 4229--4248, 2021.

\bibitem[Ait et~al.(2023)Ait, Izquierdo, and Cabot]{Ait2023}
\BIBentryALTinterwordspacing
A.~Ait, J.~L.~C. Izquierdo, and J.~Cabot, ``{HFCommunity}: {A} {Tool} to {Analyze} the {Hugging} {Face} {Hub} {Community},'' in \emph{2023 {IEEE} {International} {Conference} on {Software} {Analysis}, {Evolution} and {Reengineering} ({SANER})}.\hskip 1em plus 0.5em minus 0.4em\relax Taipa, Macao: IEEE, Mar. 2023, pp. 728--732. [Online]. Available: \url{https://ieeexplore.ieee.org/document/10123660/}
\BIBentrySTDinterwordspacing

\bibitem[Sarwar et~al.(2020)Sarwar, Zafar, Mkaouer, Walia, and Malik]{sarwar2020multi}
M.~U. Sarwar, S.~Zafar, M.~W. Mkaouer, G.~S. Walia, and M.~Z. Malik, ``Multi-label classification of commit messages using transfer learning,'' in \emph{2020 IEEE International Symposium on Software Reliability Engineering Workshops (ISSREW)}.\hskip 1em plus 0.5em minus 0.4em\relax IEEE, 2020, pp. 37--42.

\bibitem[Swanson(1976)]{swanson1976dimensions}
E.~B. Swanson, ``The dimensions of maintenance,'' in \emph{Proceedings of the 2nd international conference on Software engineering}, 1976, pp. 492--497.

\bibitem[hug()]{huggingfaceModelsHugging}
``{M}odels - {H}ugging {F}ace --- huggingface.co,'' \url{https://huggingface.co/models}.

\bibitem[Kitchenham and Pfleeger(2002)]{kitchenham2002principles}
B.~Kitchenham and S.~L. Pfleeger, ``Principles of survey research: part 5: populations and samples,'' \emph{ACM SIGSOFT Software Engineering Notes}, vol.~27, no.~5, pp. 17--20, 2002.

\bibitem[Hedges and Olkin(2014)]{hedges2014statistical}
L.~V. Hedges and I.~Olkin, \emph{Statistical methods for meta-analysis}.\hskip 1em plus 0.5em minus 0.4em\relax Academic press, 2014.

\end{thebibliography}

\end{document}